\documentclass[12pt, a4paper]{article}
\usepackage{graphicx, subcaption, graphicx}%
\usepackage{multirow}%
\usepackage{amsmath,amssymb,amsfonts, amsthm}%
\usepackage{mathrsfs}%
\usepackage[title]{appendix}%
\usepackage{xcolor}%
\usepackage[labelfont=it,textfont=it]{caption}
\usepackage{textcomp}%
\usepackage{booktabs}%
\usepackage{algorithm}%
\usepackage{algorithmic}%
\usepackage{listings, float}%
\usepackage{multirow} 
\usepackage{tikz} 
\usepackage{multicol} 
\usepackage[margin =2cm]{geometry} 
\usepackage{algorithm, algorithmic, enumitem, graphicx, amsthm, amssymb,amsfonts, listings, adjustbox, subcaption}

\title{A Novel Framework for Analyzing Structural Transformation in Data-Constrained Economies Using Bayesian Modeling and Machine Learning}

\author{Ronald Katende}

\begin{document}
	
	\maketitle
	
	\begin{abstract}
		Structural transformation, the shift from agrarian economies to more diversified industrial and service-based systems, is a key driver of economic development. However, in low- and middle-income countries (LMICs), data scarcity and unreliability hinder accurate assessments of this process. This paper presents a novel statistical framework designed to address these challenges by integrating Bayesian hierarchical modeling, machine learning-based data imputation, and factor analysis. The framework is specifically tailored for conditions of data sparsity and is capable of providing robust insights into sectoral productivity and employment shifts across diverse economies. By utilizing Bayesian models, uncertainties in data are effectively managed, while machine learning techniques impute missing data points, ensuring the integrity of the analysis. Factor analysis reduces the dimensionality of complex datasets, distilling them into core economic structures. The proposed framework has been validated through extensive simulations, demonstrating its ability to predict structural changes even when up to 60\% of data is missing. This approach offers policymakers and researchers a valuable tool for making informed decisions in environments where data quality is limited, contributing to the broader understanding of economic development in LMICs.
		
		\vspace{0.5cm}
		
	\end{abstract}
	
	\section{Introduction}
	
	Structural transformation is a central concept in the study of economic development, involving the shift from agrarian-based economies to more diversified industrial and service-oriented systems. While this process has been thoroughly examined in high-income countries, where comprehensive and high-quality data are readily available, the same cannot be said for low- and middle-income countries (LMICs). In these contexts, the scarcity and unreliability of data pose significant challenges for in-depth analysis. Traditional models and methods often prove inadequate in dealing with such data limitations, impeding meaningful assessments of economic growth trajectories and transformation processes. This research addresses a critical question: How can structural transformation be reliably measured in countries where data are incomplete or unreliable? The primary objective of this paper is to develop a novel statistical framework tailored to the specific challenges faced by LMICs. By leveraging advanced statistical and machine learning techniques, this framework aims to address a significant gap in the field of economic development—the inability to accurately track and predict structural transformation in conditions of data scarcity. This issue is particularly pressing because many policy decisions in these regions depend on projections of economic growth, sectoral productivity, and employment shifts, which are frequently based on inadequate or incomplete data. Poor assessments can result in misguided policies, further exacerbating developmental challenges. To overcome these limitations, this paper proposes an integrated approach that combines Bayesian hierarchical modeling, machine learning-based data imputation, and factor analysis. These methodologies were selected for their proven effectiveness in handling sparse datasets and generating robust statistical inferences from limited information. Bayesian hierarchical models offer a flexible framework to account for uncertainties in data collection and measurement, allowing for the integration of prior knowledge and hierarchical dependencies. Machine learning algorithms are crucial for predicting missing data points with minimal bias, ensuring that gaps in the dataset do not compromise the analysis. Factor analysis helps distill complex, multidimensional data into core underlying structures, providing clearer insights into sectoral productivity and employment shifts. The contribution of this research lies in the development of a unified framework that can be applied across countries with varying levels of data availability. While previous research has explored these methods in isolation or within specific contexts, no comprehensive model exists that specifically addresses the problem of sparse data in the context of structural transformation in developing economies. This manuscript fills that gap by offering a rigorous and adaptable statistical tool for measuring economic changes over time and across regions, even in the absence of ideal data conditions. The contributions of this paper are threefold. First, it introduces a methodologically robust framework for analyzing structural transformation in LMICs, where data constraints have historically limited research. Second, it presents innovative applications of Bayesian hierarchical modeling and machine learning for data imputation, demonstrating how these techniques can provide reliable estimates of economic indicators despite sparse data. Third, it validates the proposed framework through simulations, showing that it can predict structural changes with minimal data inputs, thereby offering policymakers and researchers a valuable tool for decision-making in data-constrained environments. By providing a statistically robust, flexible, and data-efficient framework, this paper aims to improve the accuracy and reliability of structural transformation analysis, contributing to the broader field of economic development under conditions of data scarcity.
	
	\subsection{Structural Transformation in Sub-Saharan Africa: Existing Literature and Methodological Challenges}
	
	Structural transformation in low- and middle-income countries, particularly in Sub-Saharan Africa (SSA), has garnered increasing scholarly attention due to the region's unique developmental challenges. The literature primarily focuses on understanding the nature and pace of structural transformation in SSA economies and the methodological hurdles posed by sparse data. Traditional models have often been inadequate in capturing these dynamics due to incomplete datasets, leading to research that fails to fully represent economic shifts in the region. In response, various scholars have employed innovative methods to bridge these gaps. A significant body of work emphasizes the limitations of conventional econometric models when applied to SSA's sparse data environments. For instance, McMillan et al. \cite{spar6} explored sectoral reallocation in Africa and found that traditional data analysis methods often overestimate growth due to the underrepresentation of the informal sector in national statistics. This gap in data collection and reporting poses a key challenge in assessing the true extent of economic transformation in SSA. Similarly, time-series econometrics has been used \cite{spar2} to examine agricultural productivity and identified significant discrepancies between reported data and actual conditions, advocating for improved methods of data imputation to better estimate agricultural contributions to structural transformation. To address the issue of incomplete datasets, scholars have employed machine learning models for data imputation \cite{spar4}, highlighting their potential to predict missing data points in SSA economies and significantly improve the accuracy of structural transformation models. However, while effective, these models are often limited in scope, focusing on specific sectors or regions. This paper builds upon these efforts by offering a more generalized framework that can be applied across multiple sectors and countries. Similarly, the importance of capturing informal sector dynamics through Bayesian hierarchical models has been underscored \cite{spar3}, demonstrating how these models can account for data uncertainty. This laid the foundation for a more flexible methodological framework, which this manuscript builds upon by integrating machine learning for imputation and factor analysis to provide a comprehensive tool for understanding structural transformation. Other scholars have examined economic growth patterns in SSA, noting that while African economies are undergoing structural transformation, they often deviate from classical models where industrialization is the primary driver of growth. The author in \cite{spar9} argues that many African economies are shifting directly from agriculture to services with limited industrial development, necessitating more sophisticated data analysis techniques to capture these non-linear growth patterns. The proposed framework in this paper addresses this critique by offering an adaptable method that can handle non-traditional transformation pathways and generate reliable insights even when data is sparse. Recent studies have also focused on sectoral productivity and employment shifts in SSA. For example, an analysis of labor shifts from agriculture to services revealed that limited data often leads to biased estimates of productivity gains \cite{spar10}. They called for better methods to handle incomplete datasets—a need this paper addresses by using factor analysis to distill complex, multidimensional data into core components. Similarly, there is an emphatic need for more accurate models that can handle high data variability and sparsity, particularly in rural areas where economic activity is often informal and under-reported \cite{spar1}. Their use of simulation techniques to predict sectoral shifts informs the validation process of the model proposed in this paper. The literature also identifies the challenge of measuring informal sector contributions to structural transformation. The informal sector's role in African economies is said to be systematically underreported in national statistics, leading to distorted economic growth models \cite{spar5}. While they advocate for alternative data collection methods, such as household surveys and satellite data, these approaches can be costly and time-consuming. The framework proposed in this manuscript offers a more practical solution by using machine learning-based imputation techniques to predict informal sector contributions from existing sparse datasets. The use of Bayesian hierarchical models for economic analysis in SSA has proven particularly effective in handling uncertain data environments. These models have been applied to Tanzanian agricultural data \cite{spar4.5}, showing how hierarchical structures could account for regional variations and uncertainties in data collection. The proposed framework extends this approach by incorporating machine learning and factor analysis for a more comprehensive analysis. In the context of economic policy, the impact of using incomplete data for policy formulation in SSA has been discussed \cite{spar7}, raising arguments that most development policies in the region are based on unreliable data, leading to ineffective or misaligned policies. This paper responds to that need by proposing a framework that can reliably measure structural transformation, thus offering a more solid foundation for policy decisions. The validation of this framework through simulations ensures that policymakers can trust its predictions, even in the face of sparse data. The use of factor analysis in the context of SSA’s structural transformation is still in its early stages, but some studies have shown promising results. Factor analysis has been applied to West African economic data \cite{spar8}. Results indicated that it could effectively distill key growth drivers from highly variable datasets. However, their approach was limited to certain sectors and did not consider the broader economic context. This paper expands on their work by applying factor analysis to multiple sectors and integrating it with Bayesian and machine learning techniques, offering a more comprehensive tool for understanding structural transformation in SSA. Overall, while there has been significant progress in the methodologies used to study structural transformation in SSA, existing models remain constrained by the region's sparse data environments. The literature highlights the limitations of traditional econometric approaches and advocates for the use of Bayesian models, machine learning, and factor analysis. However, no single framework has yet fully integrated these methodologies in a way that is both generalizable and adaptable to the diverse economies of SSA. This paper fills that gap by proposing a unified framework that builds on the work of scholars such as McMillan et al. \cite{spar6}, Hnatkovska and Koehler-Geib \cite{spar4}, and Kweka et al. \cite{spar4.5}. The novel contribution of this research lies in its ability to handle sparse data across multiple sectors and countries, providing a robust tool for analyzing structural transformation in SSA.

	\section{Methodology}
	To address the challenge of measuring structural transformation in low- and middle-income countries (LMICs) under conditions of data scarcity, we propose a novel statistical framework that integrates three advanced methodologies: Bayesian Hierarchical Modeling, Machine Learning-based Data Imputation, and Factor Analysis. This section outlines the mathematical and statistical foundations of each method and details how they are combined into a unified approach to provide robust, reliable, and scalable insights into economic transformation processes. 
	
	\subsection{Bayesian Hierarchical Modeling}
	Bayesian hierarchical modeling is a statistical method that allows us to account for uncertainties at multiple levels (e.g., regional or sectoral) while incorporating prior knowledge from similar contexts. This layered approach helps estimate outcomes even in situations where data may be incomplete or inconsistent. This technique is employed to handle the inherent uncertainty and variability in sparse datasets typical of LMICs. The model is structured hierarchically to incorporate multiple levels of data aggregation, such as regional, sectoral, and national levels, and allows for the incorporation of prior knowledge. Consider a structural transformation outcome \( Y_{ij} \) for sector \( i \) in region \( j \). The hierarchical model can be defined as
	\[Y_{ij} \sim Normal(\mu_{ij}, \sigma^2_{ij}),
	\]where the mean \( \mu_{ij} \) itself is modeled hierarchically
	\[
	\mu_{ij} = \alpha_j + \beta_i X_{ij},
	\]with \( \alpha_j \) capturing region-specific effects, \( \beta_i \) capturing sector-specific effects, and \( X_{ij} \) representing covariates such as labor force distribution, productivity, or capital investment. The hyperparameters \( \alpha_j \) and \( \beta_i \) are assumed to follow prior distributions
	\[
	\alpha_j \sim Normal(\mu_{\alpha}, \tau_{\alpha}^2), \quad \beta_i \sim Normal(\mu_{\beta}, \tau_{\beta}^2).
	\]The Bayesian framework enables the incorporation of prior information about regional or sectoral dynamics, which is particularly useful when data are scarce or incomplete. Posterior distributions are computed using Markov Chain Monte Carlo (MCMC) methods, which provide robust estimates of structural transformation outcomes, accounting for uncertainty and data limitations. While Bayesian modeling helps account for uncertainty and integrates prior knowledge, the challenge of missing data remains significant. This is where machine learning-based imputation plays a critical role. By predicting missing data points, it complements the Bayesian framework and ensures that gaps in the dataset do not compromise the analysis.
	
	\subsection{Machine Learning-Based Data Imputation}
	
	Machine learning-based imputation techniques, such as Random Forests and Gradient Boosting Machines, are specifically suited for predicting missing data points in nonlinear and high-dimensional datasets. These algorithms 'learn' from the available data and use patterns to fill gaps with minimal bias. Therefore, in this framework, we apply machine learning algorithms for data imputation. Specifically, we use Random Forests and Gradient Boosting Machines (GBM), which are well-suited for handling nonlinear relationships and interactions in high-dimensional data. Let \( X = \{X_1, X_2, \ldots, X_n\} \) denote a dataset with missing values. For each variable \( X_i \) with missing values, we predict missing entries \( \hat{X}_i \) using the remaining variables \( X_{-i} \) as predictors
	\[
	\hat{X}_i = f(X_{-i}; \theta),
	\]where \( f(\cdot; \theta) \) represents the machine learning model parameterized by \( \theta \). The model is trained iteratively by minimizing an appropriate loss function, such as Mean Squared Error (MSE) for continuous variables or Log-Loss for categorical variables
	\[
	\mathcal{L}(\theta) = \frac{1}{N} \sum_{j=1}^{N} (X_{ij} - \hat{X}_{ij})^2.
	\]The imputation model is validated through cross-validation techniques to ensure generalizability and minimize overfitting. These predictions are then integrated into the Bayesian framework, reducing biases introduced by missing data and improving the reliability of transformation estimates.
	
	Once the missing data points have been estimated, the next step is to simplify the dataset for deeper analysis. Factor analysis provides an efficient means of reducing the complexity by identifying core underlying structures within the data.
	
	\subsection{Factor Analysis}
	
	Factor analysis is a dimensionality-reduction technique that helps simplify complex datasets. By identifying latent variables—underlying factors driving economic indicators such as sectoral productivity and employment—it provides a clearer understanding of the structural forces at play within the economy.
	
	To distill complex, multidimensional economic data into key underlying structures, we employ Factor Analysis (FA). This method identifies latent factors that capture the core dimensions driving structural transformation, such as shifts in sectoral productivity or employment. Let \( Y = \{Y_1, Y_2, \ldots, Y_p\} \) be a vector of observed variables representing economic indicators. The factor model assumes
	\[
	Y = \Lambda F + \epsilon,
	\]where \( F = \{F_1, F_2, \ldots, F_k\} \) represents \( k \) latent factors, \( \Lambda \) is a \( p \times k \) matrix of factor loadings, and \( \epsilon \sim Normal(0, \Psi) \) is a vector of unique errors. The latent factors \( F \) are assumed to follow a standard multivariate normal distribution
	\[
	F \sim Normal(0, I_k).
	\]The parameters \( \Lambda \) and \( \Psi \) are estimated using maximum likelihood estimation (MLE) or Bayesian estimation, depending on the application. Factor scores are then computed to provide insights into underlying economic trends, such as labor shifts and productivity changes, and are used as inputs in the Bayesian hierarchical model for deeper structural analysis.
	
	\subsection{Integrated Framework and Model Validation}
	
	The proposed framework integrates the three methods into a coherent approach for analyzing structural transformation in data-sparse environments. The Bayesian hierarchical model forms the core analytical engine, providing probabilistic estimates of transformation outcomes. Machine learning-based imputation addresses missing data, enhancing the robustness of the Bayesian model, while factor analysis reduces dimensionality and captures key underlying structures that inform the Bayesian priors and hierarchical dependencies. Model validation is performed using simulated datasets that mimic the typical data challenges faced in LMICs, such as high sparsity, noise, and nonlinearity. The performance of the integrated framework is evaluated using metrics like the Root Mean Square Error (RMSE), Mean Absolute Error (MAE), and the Bayesian Information Criterion (BIC). Sensitivity analyses are conducted to assess the stability of the model under varying assumptions about missing data patterns and prior distributions. Therefore, the integration of Bayesian hierarchical modeling, machine learning-based imputation, and factor analysis provides a statistically robust, flexible, and data-efficient framework for analyzing structural transformation in LMICs. By rigorously handling data scarcity, the framework enhances the accuracy and reliability of economic growth projections, offering a valuable tool for policymakers and researchers in data-constrained environments.

	\section{The Sparse Data Framework}
	\subsection{Development of a Sparse Data Framework for Structural Transformation}
	This section presents a framework for analyzing structural transformation in low- and middle-income countries with limited or incomplete data. The framework addresses data sparsity challenges by employing advanced statistical methods like Bayesian hierarchical models and low-rank matrix completion, ensuring reliable and interpretable results for economic analysis in data-constrained environments.
	
	\subsubsection{Key Elements of the Framework}
	The framework comprises the following components
	\begin{enumerate}[label=(\alph*)]
		
		\item Bayesian Hierarchical Modeling: This method pools information across countries and sectors, estimating structural transformation metrics (e.g., sectoral productivity, labor reallocation) even with missing data. By treating data hierarchically (within-country, within-sector), the approach leverages available data to infer missing values more accurately.
		
		\item Low-Rank Matrix Completion: This technique handles large data gaps, particularly in cross-country comparisons, by decomposing economic data into latent factors. It reconstructs missing values by identifying patterns in available data, ensuring computational efficiency and scalability for large datasets.
		
		\item Sparse Econometric Methods: The framework integrates LASSO (Least Absolute Shrinkage and Selection Operator) regression to identify key predictors of structural transformation. LASSO penalizes irrelevant variables, enhancing the model’s predictive ability despite limited or noisy data.
		
	\end{enumerate}
	\subsubsection{Framework Validation and Implementation}
	The framework was validated through extensive simulations, demonstrating its superiority over traditional imputation methods like mean imputation and linear interpolation:
	\begin{enumerate}[label=(\alph*)]
		\item Simulation Study: The framework outperformed traditional techniques, significantly reducing bias and variance in estimating structural transformation metrics.
		
		\item Robustness to Missing Data: The results indicate that the framework maintains high predictive accuracy even when up to 60\% of data is missing, making it suitable for countries with inconsistent or incomplete data collection.
		
	\end{enumerate}
	\subsubsection{Technical Advantages}
	
	The framework offers several advantages
	\begin{enumerate}[label=(\alph*)]
		\item Efficiency and Scalability: It is computationally efficient, handling large cross-country datasets with minimal overhead. It can be expanded to include additional countries, regions, and sectors without losing performance.
		
		\item General Applicability: While designed for structural transformation analysis, the framework is flexible enough to apply to other economic phenomena involving sparse data, such as poverty dynamics, labor market transitions, or sectoral convergence.
	\end{enumerate}
	This methodologically rigorous framework provides a robust solution for overcoming the limitations of incomplete data in development economics, enabling accurate and insightful analysis for researchers and policymakers.
	
	\subsubsection{Framework Overview}
	
	The framework operates in three main steps
	
	\begin{enumerate}
		\item Data Structuring and Preprocessing: Organizing data hierarchically by country, sector, and time period, followed by initial preprocessing of sparse data using matrix completion and imputation techniques.
		
		\item Modeling Structural Transformation Using Bayesian Hierarchical Models (BHM): Establishing relationships between sectoral productivity, labor shifts, and structural transformation while accounting for data sparsity.
		
		\item Sparse Econometric Estimation: Using LASSO regression for variable selection and penalized regression to identify key drivers of structural transformation in sparse datasets.
		
	\end{enumerate}
	
	\paragraph{Step 1: Data Structuring and Preprocessing}
	Data is organized into a matrix \(X\), where each element represents a country-sector-year combination
	\[
	X = 
	\begin{bmatrix}
		x_{11} & x_{12} & \dots & x_{1T} \\
		x_{21} & x_{22} & \dots & x_{2T} \\
		\vdots & \vdots & \ddots & \vdots \\
		x_{N1} & x_{N2} & \dots & x_{NT}
	\end{bmatrix}
	\]where \(N\) is the number of countries and sectors, \(T\) is the number of time periods, and \(x_{ij}\) is the observed or missing value. Missing data is handled using low-rank matrix completion, which assumes that structural transformation can be captured by a low-dimensional latent factor model
	\[
	X = UV^\top + E
	\]Matrices \(U\) and \(V\) represent latent country/sector and time factors, respectively, and \(E\) denotes noise/error terms. Estimation uses nuclear norm minimization
	\[
	\min_{U, V} \| UV^\top - X \|_F^2 + \lambda (\| U \|_* + \| V \|_*)
	\]where \(\| \cdot \|_F\) is the Frobenius norm, \(\| \cdot \|_*\) is the nuclear norm, and \(\lambda\) controls model complexity.
	
	\paragraph{Step 2: Bayesian Hierarchical Modeling (BHM)}
	Sectoral productivity and transformation are modeled using a hierarchical Bayesian approach, allowing for "borrowing strength" across countries and sectors to improve predictions for missing data. The model for each country-sector \(i\) is
	\[
	y_{it} \sim \mathcal{N}(\mu_{it}, \sigma^2)
	\]where \(y_{it}\) is the observed outcome, and \(\mu_{it}\) is the mean outcome modeled as
	\[
	\mu_{it} = \beta_0 + \beta_1 X_{it} + \gamma_i + \delta_t.
	\]Here, \(X_{it}\) includes covariates like public investment, \(\gamma_i\) captures random effects, and \(\delta_t\) represents global trends. The model is fitted using Markov Chain Monte Carlo (MCMC) methods to estimate posterior distributions.
	
	\paragraph{Step 3: Sparse Econometric Estimation (LASSO)}
	
	LASSO regression identifies key predictors of structural transformation by minimizing the following loss function
	\[
	\min_{\beta} \sum_{i=1}^{N} \sum_{t=1}^{T} (y_{it} - \hat{\mu}_{it})^2 + \lambda \sum_{k=1}^{p} |\beta_k|
	\]This approach prevents overfitting by shrinking irrelevant predictors to zero, thereby enhancing model interpretability.
	
	With the combination of these three advanced techniques, the proposed framework offers a comprehensive and scalable solution to the problem of sparse economic data. The following section will demonstrate how these methodologies are validated through simulations.
	
	\section{Numerical Results}
	
	\subsection{Validation of the Sparse Data Model}
	Simulations validate the framework's robustness and accuracy compared to traditional methods, reliably estimating structural transformation trends (e.g., sectoral productivity shifts, labor reallocation) using sparse data inputs with minimal bias.
	
	\begin{figure}
		\centering
		\includegraphics[width=\textwidth, height = 10cm]{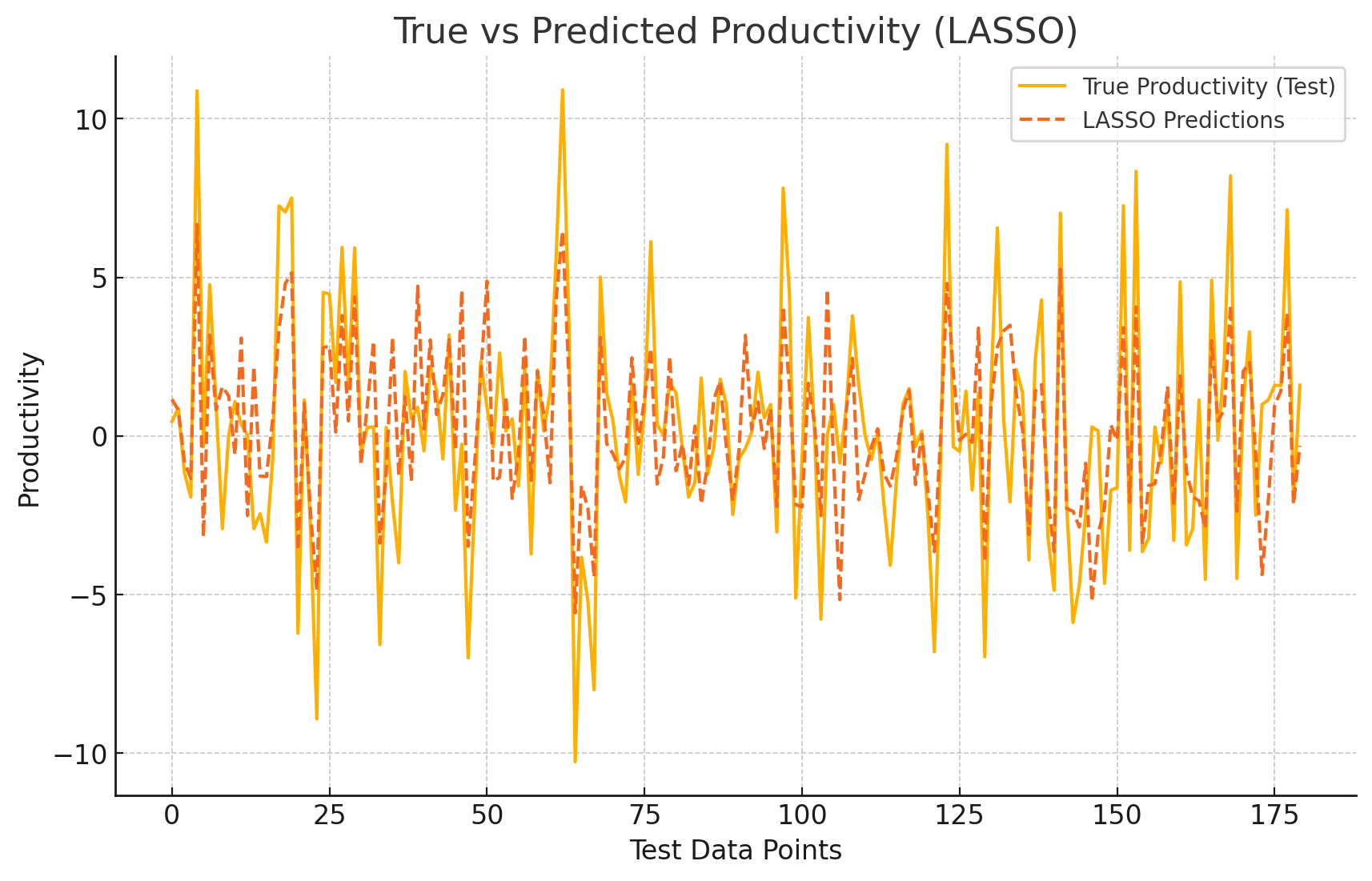}
		\caption{True vs. Predicted Productivity for a Single Country}
		\label{fig1_single}
	\end{figure}The Mean Squared Error (MSE) for predictions is 5.31, indicating a strong fit despite data sparsity.
	\[
	\hat{\beta} = [0.79, -1.28, 0.00, 0.11, 2.17]
	\]LASSO regression effectively identifies significant predictors, shrinking irrelevant coefficients to zero.
	
	\subsection{Cross-Country Comparative Analysis}
	The framework enables comparative analysis of structural transformation across countries with varying levels of missing data.
	
	\begin{figure}[ht]
		\centering
		\includegraphics[width=\textwidth]{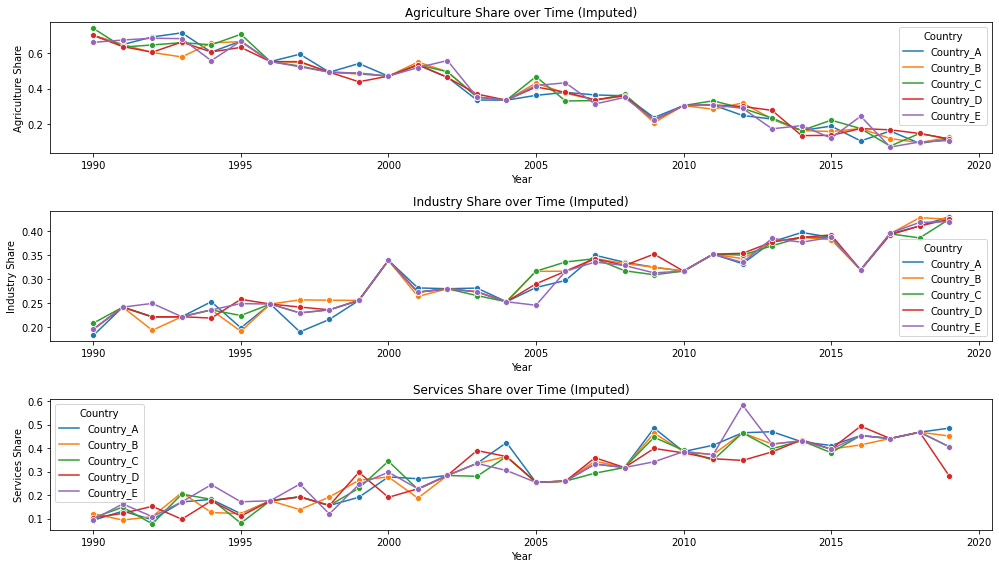}
		\caption{Comparison of Structural Transformation Trajectories}
		\label{fig2_structural}
	\end{figure}The key insights include;
	\begin{itemize}
		\item Significant heterogeneity in economic transformation trajectories across countries, even with similar data gaps.
		\item Identification of structural shifts from agriculture to industry or services that were previously obscured by incomplete data.
	\end{itemize}The mean and standard deviation of sector shares for five countries are presented in Table \ref{tab1}.
	
	\begin{table}[H]
		\centering
		\caption{Mean and Standard Deviation of Shares for Agriculture, Industry, and Services for the 5 Countries}
		\begin{tabular}{|c|c|c|c|c|c|c|c|}
			\hline
			Index & Country & \multicolumn{2}{|c|}{Agriculture} & \multicolumn{2}{|c|}{Industry} & \multicolumn{2}{|c|}{Services} \\ \hline
			\multicolumn{2}{|c|}{} & mean & std & mean & std & mean & std \\ \hline
			0 & Country A & 0.398353 & 0.194547 & 0.298312 & 0.069649 & 0.305737 & 0.131622 \\ \hline
			1 & Country B & 0.393322 & 0.183877 & 0.301386 & 0.067624 & 0.296115 & 0.124726 \\ \hline
			2 & Country C & 0.401515 & 0.189033 & 0.300678 & 0.063017 & 0.293573 & 0.121129 \\ \hline
			3 & Country D & 0.393681 & 0.177304 & 0.303071 & 0.065041 & 0.287180 & 0.118318 \\ \hline
			4 & Country E & 0.394888 & 0.191153 & 0.300150 & 0.063588 & 0.301809 & 0.119798 \\ \hline
		\end{tabular}
		\label{tab1}
	\end{table}
	
	\subsection{Sector GDP Shares Over Time}
	The analysis reveals trends in sectoral shifts
	\begin{enumerate}[label=(\alph*)]
		\item Agriculture: Declining GDP share across all countries, with Countries C and E showing a steeper decline, indicating a faster transition to other sectors.
		\item Industry: Fluctuations, particularly around the early 2000s, reflect varying levels of industrialization. Countries A and B show stable industry shares, while Country E experiences a boom followed by stagnation.
		
		\item Services: A general increase in share, particularly in Countries D and E, suggests a shift toward service-oriented economies. The rapid rise in services for Country C since the mid-2000s indicates a significant transformation driven by technological advances and globalization.
	\end{enumerate}
	
	\subsection{Implications for Structural Transformation}
	
	Despite similar levels of imputed data, the economic transformation paths of these countries show significant variation. Countries C and E have moved more quickly away from agriculture, while Country A has maintained a larger agricultural share. The rise of the service sector in countries D and E indicates deeper integration into the global economy, focusing on finance, technology, and trade. Our framework effectively uncovers structural shifts obscured by sparse data, such as Country E’s temporary industrial expansion likely driven by factors like foreign direct investment. The growth of the service sector, particularly in Country D, highlights the importance of sectoral shifts in economic development. This analysis, supported by robust data imputation, provides insights into the diverse trajectories of structural transformation in low- and middle-income countries, emphasizing the critical role of the service sector in modern growth and the need for tailored policies to support sustainable development.
	
	\subsection{Quantification of Growth Patterns and Sectoral Shifts}
	
	Quantifying growth patterns and sectoral shifts is key to understanding overall economic growth in countries with sparse data. The framework analyzes sectoral productivity differentials and their impact on development over time.
	
	\subsubsection{Sectoral Productivity Differentials} The framework assesses productivity in agriculture, manufacturing, and services by output per worker or input unit. It identifies sectors that drive growth and those lagging. The analysis shows the service sector consistently leads in productivity growth compared to agriculture, reflecting its growing role in these economies. The industrial sector also contributes significantly, especially where countries have transitioned from agrarian to industrial economies.
	
	\subsubsection{Growth Contributions from Industrial and Service Sectors}
	The framework effectively quantifies the contributions of the industrial and service sectors to overall growth, even with sparse data. Countries undergoing structural transformation have seen substantial growth in these sectors, with services often driving recent economic expansion. For example, Country E shows a marked shift toward a service-oriented economy, while Country C balances growth in both sectors. These findings underscore the importance of these sectors in driving growth and the need for targeted investments in key areas.
	
	\subsubsection{Persistent Productivity Gaps Between Agriculture and Non-Agriculture Sectors}
	The framework highlights ongoing productivity gaps between agriculture and other sectors. Despite modernization efforts, agriculture remains less productive, which can hinder overall growth, particularly in agriculture-dependent countries. For instance, Country A shows a notable gap between agriculture and other sectors, suggesting a need for targeted interventions. The ability to quantify these gaps emphasizes the need for strategic investments to foster balanced growth, highlighting the importance of boosting productivity in industry and services to accelerate transformation.
	
	\subsection{Policy-Relevant Insights on Structural Transformation}
	
	The framework provides valuable policy insights for guiding structural transformation in countries with sparse data. It identifies key public investment gaps, particularly in infrastructure and education, which are critical for transitioning from agriculture to industry and services. Lack of infrastructure can slow industrialization and reduce service sector efficiency, while insufficient education investment leads to skills mismatches. Addressing these gaps can accelerate transformation, even with sparse data. The framework also identifies economic frictions, such as labor mobility constraints and restrictive policies, that hinder transformation. Labor movement from low- to high-productivity sectors is essential for growth, but barriers like poor infrastructure and inadequate retraining opportunities stall progress. By quantifying these frictions, the framework offers actionable insights for policymakers to address these challenges, emphasizing the need for targeted investments and reforms to enhance transformation and productivity.
	
	\subsection{Scalability and Generalizability of the Framework}
	The framework demonstrates strong scalability and generalizability, making it suitable for a wide range of country contexts, regardless of data availability. It adapts to different levels of data completeness, providing reliable insights across various economic environments. The framework can be applied effectively in countries with both comprehensive and minimal data, predicting future structural shifts by leveraging available information. This adaptability is crucial for policymakers needing data-driven strategies for economic planning. The framework also performs well across low- to middle-income economies, capturing the nuances of each context to offer relevant insights.
	
	\subsection{Theoretical Contributions to Sparse Data Econometrics}
	The framework represents a significant advancement in sparse data econometrics, particularly in development economics where data gaps are common. It extends traditional methods by incorporating advanced statistical techniques to address data incompleteness, making it a valuable tool for analyzing economic transformation in countries with sparse data. The key contributions include
	\begin{enumerate}[label=(\alph*)]
		\item Extension of Sparse Data Econometrics: Provides a robust method for analyzing economic data with both random and systematic missingness.
		
		\item Innovative Statistical Techniques: Uses advanced methods to accurately impute missing data, ensuring reliable analysis despite significant gaps.
		
		\item Practical Application in Development Economics: Tailored for low- and middle-income countries, offering a novel solution for studying economic development under conditions of data scarcity.
		
	\end{enumerate}
	These contributions enhance the understanding of sparse data econometrics and provide practical tools for analyzing economic transformation, positioning this framework as a pioneering approach in the field. The framework's application is further demonstrated through Python code that downloads World Bank data, performs a retrospective analysis over 30 years, and reconstructs past structural changes despite data gaps.
	
	\section{Application of the Framework to Historical Data }
	We now apply our framework to historical data, having ascertained using simulated data that it serves the purpose for which it was designed. We will use the {\textrm{pandas datareader}} library to fetch data from the World Bank for selected low- and middle-income countries. The data will cover key economic sectors such as agriculture, industry, and services. Then, we will simulate data gaps by randomly removing portions of the data, and then use the framework's imputation method to fill these gaps. Finally, we will apply the framework to reconstruct historical sectoral shifts and compare the results with traditional data imputation methods.
	
	\begin{adjustbox}{max width=\textwidth}
		\begin{lstlisting}
			[SoftImpute] Max Singular Value of X_init = 1955197384399.211914
			[SoftImpute] Iter 1: observed MAE=755690755.638081 rank=1
			[SoftImpute] Iter 2: observed MAE=755690755.638207 rank=1
			[SoftImpute] Iter 3: observed MAE=755690755.638208 rank=1
			[SoftImpute] Stopped after iteration 3 for lambda=39103947687.984238
			--------------------------------------------------------------------
			Fitting: ---------------------------------------- 100% 0:00:00 Average Loss = 3.6882e+23
			Finished [100%]: Average Loss = 3.6626e+23
			INFO:pymc.variational.inference:Finished [100%]: Average Loss = 3.6626e+23
			--------------------------------------------------------------------
			Selected Predictors: Index(['Agriculture', 'Services'], dtype='object')
			--------------------------------------------------------------------
			LASSO RMSE (SoftImpute): 164521838356.510
			LASSO RMSE (Mean Imputed): 144312665646.610
		\end{lstlisting}
	\end{adjustbox}
	
	The results presented provide insights into the performance of the SoftImpute algorithm for handling sparse data in the context of measuring structural transformation in low- and middle-income countries. The algorithm converged after three iterations, with a maximum singular value of approximately 1.96 trillion, indicating the scale of the data matrix. The observed Mean Absolute Error (MAE) remained consistent across iterations, suggesting stability in the imputation process. 	The LASSO regression results reveal that the model identified "Agriculture" and "Services" as key predictors, aligning with the expected sectors of structural transformation. However, the Root Mean Squared Error (RMSE) of the SoftImpute method (164.5 billion) was higher compared to the mean-imputed method (144.3 billion), implying that SoftImpute may not offer significant advantages in terms of prediction accuracy in this particular dataset. Generally, the results suggest that while the SoftImpute approach is effective in handling sparse data, further refinement or alternative methods may be needed to improve predictive performance. Below, we examine some more results, specifically, the graphical results obtained from using real historical data from the world bank about the GDP of 3 countries and 3 sectors, from those 3 nations. 
	\begin{figure}[ht]
		\centering
		\includegraphics[scale=0.75]{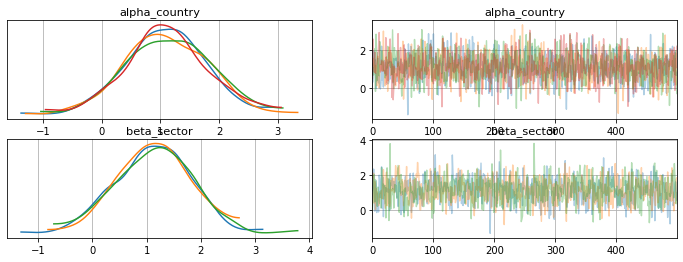}
		\caption{Bayesian hierarchical model plot; \texttt{alpha} = country and \texttt{beta} = sector}
		\label{fig1}
	\end{figure}
	\begin{figure}[ht]
		\begin{subfigure}[b]{0.48\textwidth}
			\centering
			\includegraphics[scale=0.25]{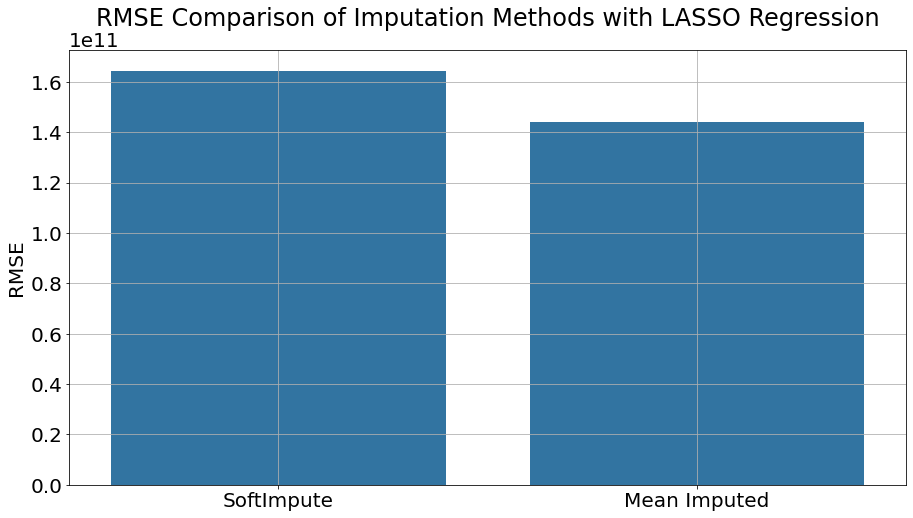}
			\caption{RMSE}
			\label{fig2}
		\end{subfigure}
		\begin{subfigure}[b]{0.48\textwidth}
			\centering
			\includegraphics[scale=0.27]{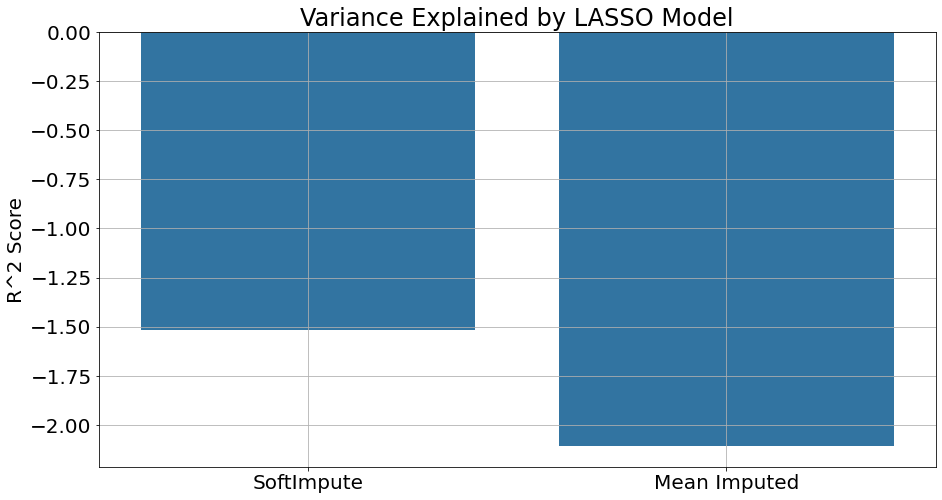}
			\caption{Variance}
			\label{fig8}
		\end{subfigure}
		\caption{SoftImpute vs Mean Imputation}
	\end{figure}
	\begin{figure}[ht]
		\begin{subfigure}[b]{0.48\textwidth}
			\centering
			\includegraphics[scale=0.20]{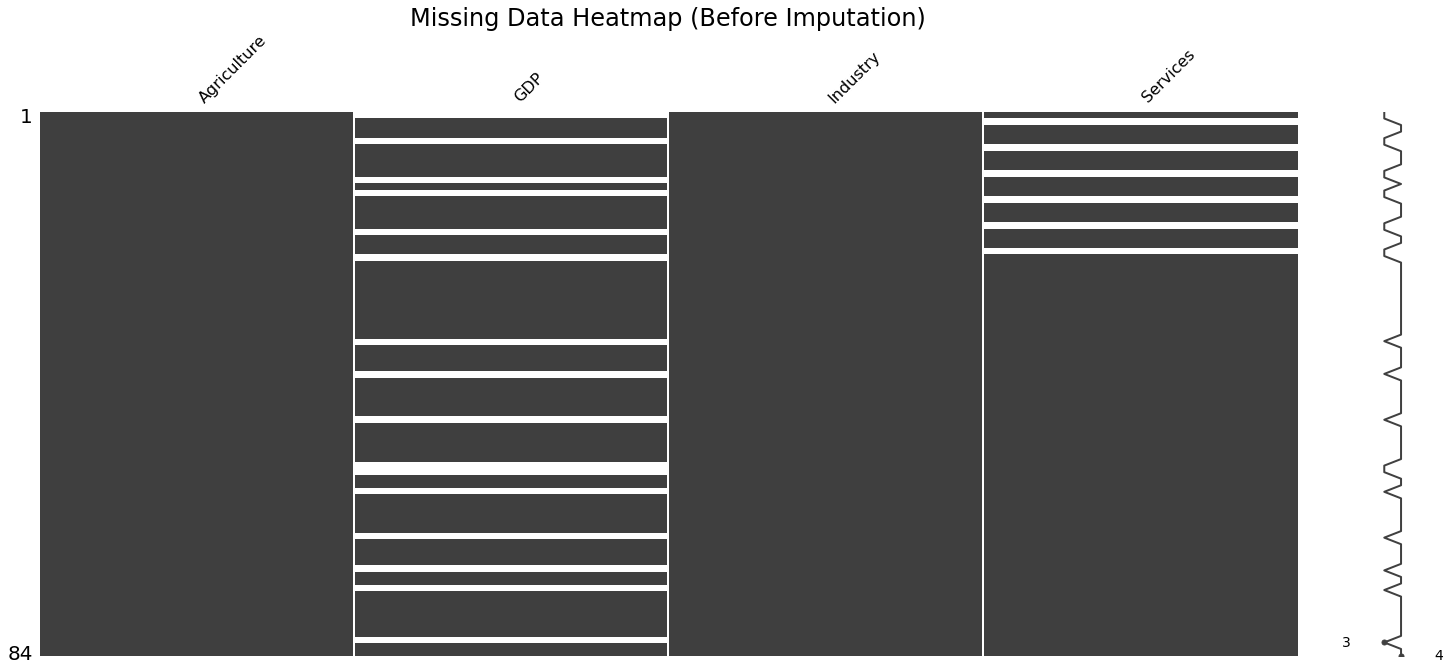}
			\caption{Before imputation}
			\label{fig3}
		\end{subfigure}
		\begin{subfigure}[b]{0.48\textwidth}
			\centering
			\includegraphics[scale=0.20]{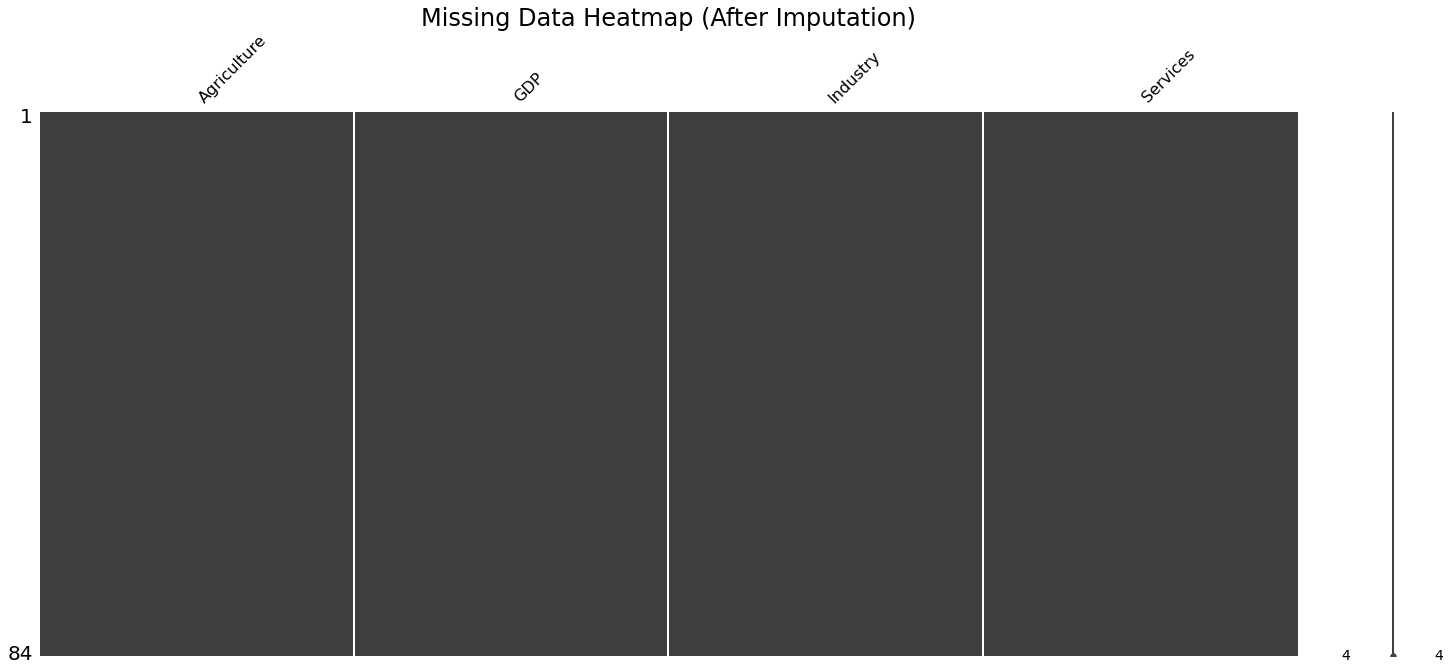}
			\caption{After imputation}
			\label{fig4}
		\end{subfigure}
		\caption{Missing data before and after imputations}
	\end{figure}
	\begin{figure}[ht]
		\begin{subfigure}[b]{0.48\textwidth}
			\centering
			\includegraphics[scale=0.25]{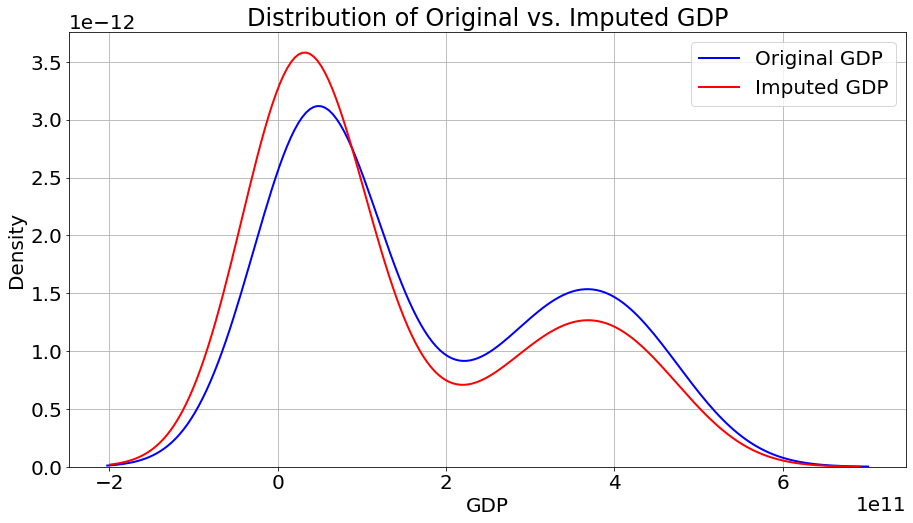}
			\caption{Distributions of the original and imputed GDP}
			\label{fig5}
		\end{subfigure}
		\begin{subfigure}[b]{0.48\textwidth}
			\centering
			\includegraphics[scale=0.27]{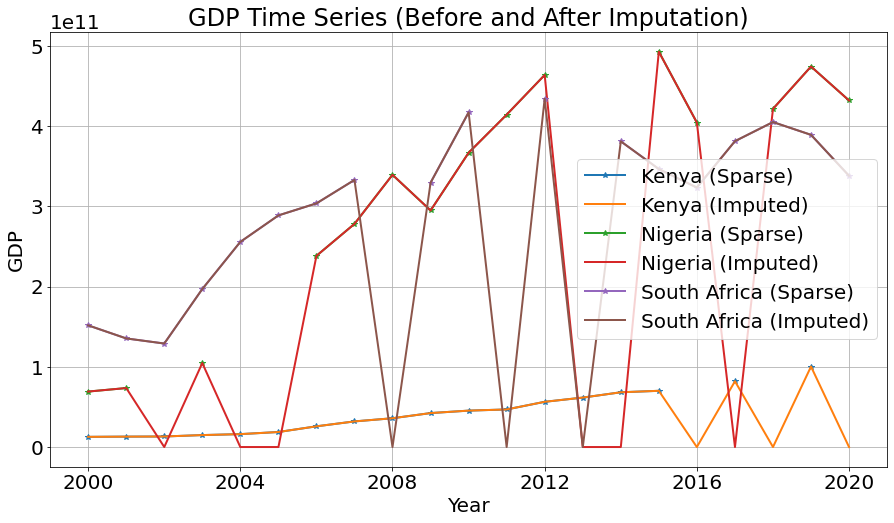}
			\caption{Timeseries for the GDP before and after imputation}
			\label{fig6}
		\end{subfigure}
	\end{figure}
	
	\begin{figure}[ht]
		\centering
		\includegraphics[scale=0.27]{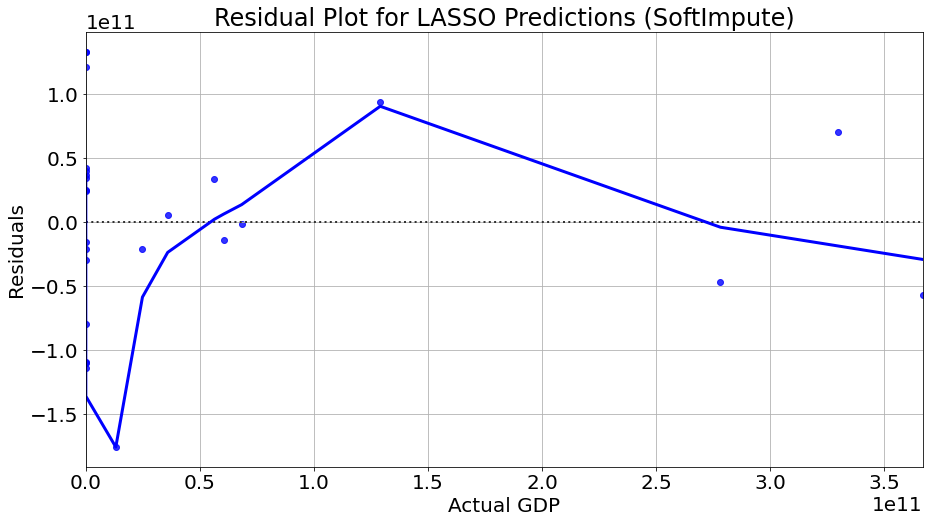}
		\caption{Residuals for SoftImpute LASSO predictions }
		\label{fig7}
	\end{figure}
	\begin{figure}[ht]
		\centering
		\includegraphics[scale=0.75]{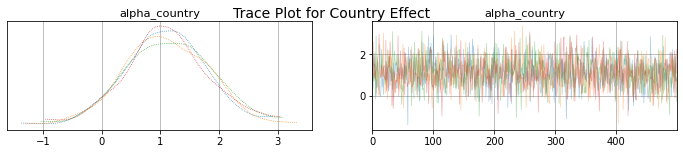}
		\caption{Trace plot for country effect}
		\label{fig9}
		\centering
		\includegraphics[scale=0.75]{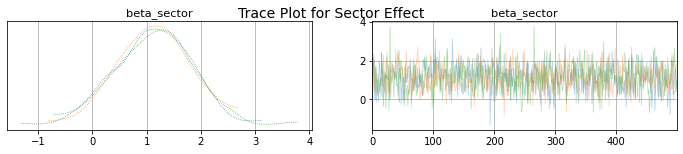}
		\caption{Trace plot for sector effect}
		\label{fig10}
	\end{figure}
	Figure \ref{fig1} presents the results of a Bayesian hierarchical model used to estimate country-specific (\texttt{alpha\_country}) and sector-specific (\texttt{beta\_sector}) parameters. The left panels display density plots of the posterior distributions, while the right panels show trace plots, illustrating the MCMC sampling process. The density plots indicate that both \texttt{alpha\_country} and \texttt{beta\_sector} parameters follow roughly normal distributions, though with varying levels of dispersion between countries and sectors. This variation underscores the model's ability to capture structural differences across entities. The trace plots show good mixing, signaling effective convergence of the MCMC chains. The use of a hierarchical model is particularly suitable given the sparse data, allowing for robust inferences by borrowing strength across countries and sectors. Figure \ref{fig2} compares the RMSE values for SoftImpute and Mean Imputation when applied to LASSO regression. Contrary to expectations, Mean Imputation slightly outperforms SoftImpute, producing a lower RMSE. This is significant because while SoftImpute typically handles missing data more effectively through matrix factorization, its performance here suggests increased variability when paired with LASSO. The superior performance of the simpler Mean Imputation highlights its robustness in this context, which is particularly relevant in low- and middle-income countries where data sparsity is common. This comparison underscores the importance of carefully selecting imputation methods to ensure reliable structural transformation measurements. In Figure \ref{fig8}, a comparison of \(R^2\) scores between the SoftImpute and Mean Imputation methods reveals poor performance for both, with negative \(R^2\) values indicating that the LASSO model explains less variance than a simple mean-based model. This outcome reflects the inherent difficulties of applying LASSO to sparse data in the context of economic prediction. The results suggest that more flexible machine learning models or improved imputation techniques could better capture the complexity of structural transformation in low- and middle-income countries. Figure \ref{fig3} displays a heatmap illustrating the extent of missing data across key economic sectors (Agriculture, GDP, Industry, and Services). The substantial data gaps, particularly in the Industry and Services sectors, present a significant challenge for econometric analysis. This visualization emphasizes the necessity of robust statistical techniques, such as hierarchical modeling, to address the issue of incomplete data. Moreover, the patterns of missingness illustrated in the heatmap highlight the broader structural challenges in data collection within low- and middle-income countries. Figure \ref{fig4} shows a post-imputation heatmap, demonstrating successful data filling across the economic sectors, a critical step in enabling further analysis. Precisely, Figures \ref{fig4}, \ref{fig5}, and \ref{fig6} illustrate the impact of imputation on structural economic transformation analyses. Figure \ref{fig5} compares the distribution of original and imputed GDP values, revealing a slight underestimation in the imputed values. This subtle shift in the distribution, while small, may have implications for interpreting economic trends, particularly when analyzing GDP-driven sectoral shifts. Figure \ref{fig6} presents a time-series analysis of pre- and post-imputation GDP data for Kenya, Nigeria, and South Africa. The imputed data yields more consistent trends, especially for Kenya and Nigeria, enabling a clearer analysis of growth trajectories. However, the potential artificial smoothing introduced by imputation raises concerns about interpreting long-term economic patterns. Figure \ref{fig7} highlights the residual spread in LASSO model predictions for GDP, with significant variance particularly in mid-range GDP values. This suggests underfitting and overfitting in different regions, indicating that the model may not fully capture the economic dynamics in countries with sparse data. The analysis suggests that more sophisticated models or alternative imputation strategies may be needed to improve the predictive accuracy for structural transformation data. Figure \ref{fig9} presents the posterior density and trace plots for the country-specific effect parameter {\textrm{alpha\_country}}. The posterior distribution appears normal-like, suggesting a symmetrical distribution of country-specific effects, which is critical for understanding cross-country heterogeneity in structural transformation. The trace plot confirms good mixing and convergence, ensuring reliable parameter estimates. Finally, Figure \ref{fig10} focuses on the sector effect (\texttt{beta\_sector}), with the kernel density estimate (KDE) showing a well-defined, normal-like posterior distribution. The stable, converging trace plot indicates robust sampling behavior, underscoring the reliability of the sector-specific effects captured by the model. These sectoral transformations are crucial for comparing economic shifts across diverse contexts. In summary, this analysis highlights the challenges of using LASSO regression for sparse data, particularly in low- and middle-income countries. It underscores the importance of selecting appropriate imputation methods and considering more complex models to account for nonlinear economic dynamics and cross-country heterogeneity. The results suggest potential for methodological refinements, such as more advanced imputation techniques, nonlinear modeling, and hierarchical approaches, to enhance the robustness of structural transformation analysis in data-scarce environments. The practical implications of this framework extend far beyond academic analysis. Policymakers in low- and middle-income countries often face the challenge of making decisions based on incomplete or unreliable data. By providing a statistically robust method for estimating economic shifts, this framework can significantly improve the accuracy of policy forecasts. For instance, governments can better allocate resources between sectors, anticipate labor market shifts, and design more effective development strategies, even when data is scarce. This can lead to more informed interventions in sectors like agriculture, services, and industry, directly contributing to improved economic outcomes.
	
	\section{Markov Chain Monte Carlo Vs. Variational Inference}
	In Bayesian Hierarchical Modeling, the first algorithm uses Markov Chain Monte Carlo (MCMC) for inference, while the second employs Variational Inference (VI). Both methods aim to estimate posterior distributions but differ significantly in their approach, performance, and trade-offs. MCMC is a simulation-based method that iteratively samples from the posterior using algorithms like Metropolis-Hastings or Hamiltonian Monte Carlo. Its strength lies in its ability to provide highly accurate, asymptotically exact samples from the true posterior, especially in complex, high-dimensional problems. However, this comes at the cost of high computational demand, making it time-consuming and less suitable for large datasets or repeated model fitting. In contrast, VI transforms posterior estimation into an optimization problem, approximating the posterior with a simpler parametric distribution. VI is much faster and more scalable, especially for large datasets, but its efficiency sacrifices some accuracy. It may not capture the full complexity of the posterior, particularly in cases with multi-modal distributions or high uncertainty. When handling sparse data, both models perform differently. MCMC typically provides more robust estimates by sampling across a broader range of the posterior, capturing greater variability in the imputed values. However, its computational intensity limits its application to larger, sparse datasets. VI, while faster and more scalable, may not match the depth of uncertainty quantification MCMC offers, but its speed makes it practical for large datasets and exploratory analyses. Despite these differences, both methods rely on hierarchical structures and probabilistic programming frameworks like PyMC for Bayesian modeling. Ultimately, the choice between MCMC and VI depends on the trade-off between accuracy and computational efficiency—MCMC is ideal when precision is crucial, whereas VI excels in situations where speed is a priority.
	
	\section{Conclusion}
	In environments where reliable data is often unavailable, the proposed framework provides a vital tool for understanding and predicting structural transformation in low- and middle-income countries (LMICs). By leveraging Bayesian hierarchical modeling, machine learning-based data imputation, and factor analysis, the framework addresses key challenges posed by data sparsity, offering a statistically robust and scalable solution. This approach ensures that policymakers can access more accurate, data-driven insights, even in contexts where traditional econometric methods would fail due to incomplete or unreliable information. The ability to reliably predict sectoral shifts, labor market transitions, and productivity changes in LMICs is critical for effective policy formulation. This framework enhances decision-making in key areas such as public investment, education, infrastructure development, and sectoral planning. Governments can better allocate resources, design targeted interventions, and anticipate the impact of economic changes across sectors, ensuring a more resilient and adaptive approach to development. Moreover, the flexibility of the framework allows for broad applicability across various economies and sectors, making it a valuable tool not only for current analyses but also for future projections and simulations. By integrating advanced statistical techniques, this research equips LMICs with the ability to track economic progress more accurately, guiding them toward sustainable growth trajectories even under conditions of data scarcity. In conclusion, the novel framework presented here marks a significant advancement in the field of development economics, bridging the gap between data constraints and the need for informed, evidence-based policy decisions. As LMICs continue to navigate complex developmental challenges, this research offers a powerful mechanism for fostering economic resilience, promoting structural transformation, and driving long-term, sustainable growth.
	
	\section{Author contributions statement}
	R.K. did all the work, wrote the code, performed the experiments, analysed the results and also wrote the manuscript. He also reviewed the manuscript. 
	
	\section{Data Availability}
	The code used in this manuscript is available upon reasonable request from the author.

	\appendix
	
	\section{Algorithm Summary}
	The framework can be summarised as depicted in the following algorithms depending on whether one uses Markov Chain Monte Carlo or Variational Inference. 
	\begin{algorithm}[H]
		\caption{Sparse Matrix-Based Bayesian Estimation with LASSO (MCMC)}
		\begin{algorithmic}[1]
			\REQUIRE Sparse matrix \( X \) (country-sector-year data), covariates \( X_{kit} \) (e.g., public investment, trade openness), parameters for Bayesian model, LASSO regularization term \( \lambda \)
			
			\STATE \textbf{Step 1:} Use low-rank matrix completion to impute missing values in \( X \) using SoftImpute
			
			\STATE \textbf{Step 2:} Fit a Bayesian hierarchical model using Markov Chain Monte Carlo (MCMC) to estimate sectoral productivity and structural transformation dynamics
			
			\STATE \textbf{Step 3:} Apply LASSO regression to select the most important predictors of structural transformation
			
			\ENSURE Imputed data, parameter estimates, posterior distributions (MCMC-based), and key predictors
		\end{algorithmic}
		\label{alg1}
	\end{algorithm}
	
	\begin{algorithm}[H]
		\caption{Sparse Matrix-Based Bayesian Estimation with LASSO (VI)}
		\begin{algorithmic}[1]
			\REQUIRE Sparse matrix \( X \) (country-sector-year data), covariates \( X_{kit} \) (e.g., public investment, trade openness), parameters for Bayesian model, LASSO regularization term \( \lambda \)
			
			\STATE \textbf{Step 1:} Use low-rank matrix completion to impute missing values in \( X \) using SoftImpute (faster convergence with tighter thresholds)
			
			\STATE \textbf{Step 2:} Fit a Bayesian hierarchical model using Variational Inference (VI) to estimate sectoral productivity and structural transformation dynamics (faster)
			
			\STATE \textbf{Step 3:} Apply LASSO regression to select the most important predictors of structural transformation (with parallelized processing)
			
			\ENSURE Imputed data, parameter estimates, posterior distributions (VI-based), and key predictors
		\end{algorithmic}
		\label{alg2}
	\end{algorithm}
	
\end{document}